\documentclass[amsmath,twocolumn,superscriptaddress,prl,aps]{revtex4}
\usepackage{amsthm,amsfonts,graphicx,verbatim,color}
\usepackage{epstopdf}
\usepackage{subfigure}
\usepackage{amsbsy} 
\usepackage{url}
\usepackage{hyperref}

\newcommand{\be}{\begin{equation}}
\newcommand{\ee}{\end{equation}}
\newcommand{\bea}{\begin{eqnarray}}
\newcommand{\eea}{\end{eqnarray}}

\renewcommand{\phi}{\varphi}
\renewcommand{\epsilon}{\varepsilon}
\renewcommand{\vec}[1]{{\bf #1}}

\renewcommand{\cite}[1]{[\onlinecite{#1}]}

\begin{document}
\title{Flat bands with local Berry curvature in multilayer graphene}
\author{Akshay Kumar}
\affiliation{Department of Physics, Princeton
University, Princeton, NJ 08544, USA}
\author{Rahul Nandkishore}
\email{rahuln@princeton.edu}
\affiliation{Princeton Center for Theoretical Science, 
Princeton University, Princeton, NJ 08544, USA}

\begin{abstract}
We demonstrate that flat bands with local Berry curvature arise naturally in chiral (ABC) multilayer graphene placed on a boron nitride (BN) substrate. 
The degree of flatness can be tuned by varying the number of graphene layers $N$. For $N=7$ the bands become nearly flat, with a small bandwidth $\sim 3.6$ meV. The two nearly flat bands coming from the $K$ and $K'$ valleys cross along lines in the reduced zone. Weak intervalley tunneling turns the bandcrossing into an avoided crossing, producing two nearly flat bands with global Chern number zero, but with local Berry curvature. The flatness of the bands suggests that many body effects will dominate the physics, while the local Berry curvature of the bands endows the system with a nontrivial quantum geometry. The quantum geometry effects manifest themselves through the quantum distance (Fubini-Study) metric, rather than the more conventional Chern number. Multilayer graphene on BN thus provides a platform for investigating the effect of interactions in a system with a non-trivial quantum distance metric, without the complication of non-zero Chern numbers. We note in passing that flat bands with non-zero Chern number can also be realized by making use of magnetic adatoms, and explicitly breaking time reversal symmetry. 
\end{abstract}
\maketitle

The behavior of strongly interacting electrons and the effect of quantum geometry are two of the most exciting fields of research in modern condensed matter physics. Strong interactions can give rise to effects such as fractionalization \cite{Rajaraman, Laughlin, Yacoby}, where the elementary excitations carry some fraction of the `fundamental' electronic quantum numbers. One popular way to access strong correlation physics is to consider band structures that are flat (non-dispersive), since in such band structures interactions naturally dominate over kinetic energy. Meanwhile, the two ingredients of quantum geometry are integrated Berry curvature (a.k.a. Chern number for two dimensional systems) and Fubini-Study metric \cite{quantum geometry haldane}, the metric measuring the quantum distance. While non-zero integrated Berry curvature can lead to the existence of states that are protected against disorder \cite{Blok, Kane, Haldane} and to effects like quantized Hall conductance of a band insulator \cite{TKNN}, non-zero Fubini Study metric can lead to phenomena such as pseudospin conservation laws in single layer graphene \cite{quantum geometry haldane} and unusual features in the current noise spectrum of a band insulator \cite{Mudry}. 
Recently, a strong interest has emerged in flat band systems as playgrounds for investigating the interplay of strong correlation effects and non-trivial quantum geometry. However, attention has been mostly focused on systems with non-zero integrated Berry curvature, \cite{Neupert,Tang,Parameswaran,Shankar,Bernevig,Regnault,Qi, Sun, Ran, Trescher}, while the effect of the Fubini-Study metric has largely been ignored. 


The influence of quantum geometry on the strong correlation physics stems from the fact that the interactions couple to electron density, and the electron density operators projected onto the flat band obey a non-trivial commutation relation if the band has a non-trivial quantum geometry \cite{Parameswaran, Rahul roy}. The general commutation relation for projected densities in a band with non-uniform Berry curvature is \cite{Rahul roy}:
\begin{equation}
[\bar{\rho}_{\vec{q1}},\bar{\rho}_{\vec{q2}}] \approx  i \vec{q_1} \wedge \vec{q_2} \sum_{\vec{k},b} [B(\vec{k}) u^{*}_{b}(\vec{k}_{+}) u_{b}(\vec{k}_{-}) \gamma_{\vec{k}_{+}}^{\dagger} \gamma_{\vec{k}_{-}}] \label{eq: FS}
\end{equation}
where $\vec{q1} \wedge \vec{q2} = \hat{z} . (\vec{q_1} \times \vec{q_2})$, $\vec{k}_{\pm}=\vec{k} \pm \frac{\vec{q_1}+\vec{q_2}}{2}$, $B(\vec{k})$ is the local Berry curvature for the band of interest for the single-particle hamiltonian $H=\sum_{\vec{k},a,b} c_{\vec{k},a}^{\dagger} h_{ab}(\vec{k}) c_{\vec{k},b}$  and the corresponding eigenstate is given by $\left | \vec{k} \right \rangle = \gamma_{\vec{k}}^{\dagger}\left | 0 \right \rangle = \sum_{a} u_{a}(\vec{k}) c_{\vec{k},a}^{\dagger} \left | 0 \right \rangle$.  Thus far, attention has been focused on systems where the Berry curvature is nearly uniform \cite{Neupert,Tang,Parameswaran,Shankar,Bernevig,Regnault,Qi, Sun, Ran, Trescher}. However, it is apparent from (\ref{eq: FS}) that non-trivial quantum geometry effects do not require a uniform Berry curvature. The influence of the quantum distance metric may be most clearly revealed in a flat band system with local Berry curvature, but zero integrated Berry curvature, since here the non-trivial quantum geometry effects (encoded e.g. in the projected density commutator) arise purely due to the metric. Thus far, theoretical studies have largely ignored this exciting direction of research, in part because of the lack of experimental realizations. 

In this Letter, we show that flat bands with local Berry curvature (but vanishing integrated Berry curvature) arise naturally in chiral multilayer graphene. Our proposal exploits the fact that ABC stacked multilayer graphene in the presence of a perpendicular electric field has a bandstructure with flat pockets that possess Berry curvature. Placing the graphene on a hexagonal Boron Nitride (BN) substrate then produces a superlattice potential such that the reduced Brillouin zone lies entirely within the flat pocket. Umklapp scattering opens  a bandgap at the reduced zone edge. 
For a $N$ layer system with $N>5$ layers, the lowest band is nearly flat, with a bandwidth $\sim 5 meV$. We have verified that this nearly flat band has a non-vanishing local Berry curvature. Chiral multilayer graphene thus provides an ideal platform for investigating the interplay of strong correlations and quantum geometry, with the quantum geometry effects coming from the (hitherto neglected) channel of the Fubuni-Study metric, rather than the more conventional channel of non-zero Chern number. As such, we believe chiral multilayer graphene offers an exciting new frontier in the study of systems with an interplay of strong correlations and quantum geometry. 

%
%
We note in passing that a non-zero integrated Berry curvature can 
 also be obtained 
by using magnetic adatoms rather than vertical electric field to open a gap between conduction and valence bands. 

\emph{ABC stacked graphene}: A single graphene sheet consists of a honeycomb lattice of carbon atoms. The honeycomb lattice consists of two sublattices, $A$ and $B$. Chiral multilayer graphene consists of graphene sheets with an $ABC$ stacking order (each succeeding sheet is rotated by $2\pi/3$ relative to the preceding sheet). Every lattice site in the bulk is either directly above or directly below another lattice site. In the $(\psi_{1A\vec{k}} \psi_{1B\vec{k}} \psi_{2A\vec{k}} \psi_{2B\vec{k}} \cdots \cdots \psi_{NA\vec{k}} \psi_{NB\vec{k}})$ basis, the nearest neighbor tight binding Hamiltonian for an $N$ layer system takes the form \cite{Zhang}
\begin{equation}
H_0 = \left(\begin{array}{ccccccc}
0& t_{\vec{p}} & 0 & 0 & 0 & 0 & \cdots \\
t^*_{\vec{p}} & 0 & \gamma & 0 & 0 & 0 & \cdots \\
0 & \gamma & 0 & t_{\vec{p}} & 0 & 0 & \cdots \\
0 & 0 & t^*_{\vec{p}} & 0 & \gamma & 0 & \cdots \\
0 & 0 & 0 & \gamma & 0 & t_{\vec{p}} & \cdots \\
0 & 0 & 0 & 0 & t^*_{\vec{p}} & 0 & \cdots \\
\vdots & \vdots & \vdots & \vdots &\vdots & \vdots & \ddots \\
\end{array}
 \right)
\end{equation}

Here, $t_{\vec{p}} = t_0 \big(\exp( i k_x a) + 2 \exp(-i \frac{k_x a }{2}) \cos(\frac{\sqrt3 k_y a}{2})\big)$, represents nearest neighbor hopping within each graphene layer ($t_0 \approx 3 eV$), and $\gamma \approx 300 meV$ represents interlayer hopping between two sites that lie on top of each other. Here $a$ is the lattice constant of graphene.

In the absence of interlayer hopping, $\gamma = 0$, the bandstructure consists of $N$ copies of the graphene bandstructure, $E = \pm |t_{\vec{p}}|$ \cite{Wallace}. Near the two inequivalent corners of the Brillouin zone, conventionally labelled $K$ and $K'$, the function $t_{\vec{p}}$ vanishes as $t_{\vec{p}} \approx v \vec{p}_+$ and $t_{\vec{p}} \approx - v \vec{p}_-$ respectively, where $\vec{p}_{\pm} = p_x \pm i p_y$ and $v=3at_{o}/2$ is the Fermi velocity for graphene. 

We now consider interlayer hopping $\gamma \neq 0$. This interlayer hopping causes all the bulk sites to dimerize, opening up a bulk gap of size $\gamma$ at the Dirac points. On the top and bottom surfaces, there are undimerized lattice sites, which give rise to gapless surface states. The low energy single particle Hamiltonian for the surface states of an $N$ layer graphene takes the form \cite{McCann, supplement, Volovik, Volovik2}
\begin{equation}
H_K(\vec{p}) = \frac{v^N}{(-\gamma)^{N-1}} \left( \begin{array}{cc} 0 & p_+^N \\ p_-^N & 0 \end{array} \right); \quad p_{\pm} = p_x \pm i p_y
\end{equation}

Here $H_K$ is the Hamiltonian in the $K$ valley, and the Hamiltonian in the $K'$ valley is given by $H_{K'}(\vec{p}) = H^*_K(-\vec{p})$. The basis is such that $(1,0)$ is a Bloch state in the A sublattice of the top layer, and $(0,1)$ is a Bloch state in the $B$ sublattice of the bottom layer.  Only the lowest order terms have been written in each matrix element. This Hamiltonian is valid up to an energy scale $\gamma$. Note that this energy scale is independent of $N$, the number of layers of graphene. The above effective Hamiltonian consists of a single bandcrossing with non-trivial Berry phase $N \pi$, and low energy dispersion $E = \pm v^N p^N / \gamma^{N-1}$. Thus, the conduction band has a very flat pocket of a size $p\sim\gamma/v$ around the $K$ and $K'$ points, and this pocket becomes perfectly flat in the limit $N\rightarrow \infty$, corresponding to `rhombohedral graphite'. The emergence of this flat pocket has previously been discussed in \cite{Volovik, Volovik2}. 

\begin{figure}
\centering
\subfigure{\includegraphics[width=0.9\columnwidth]{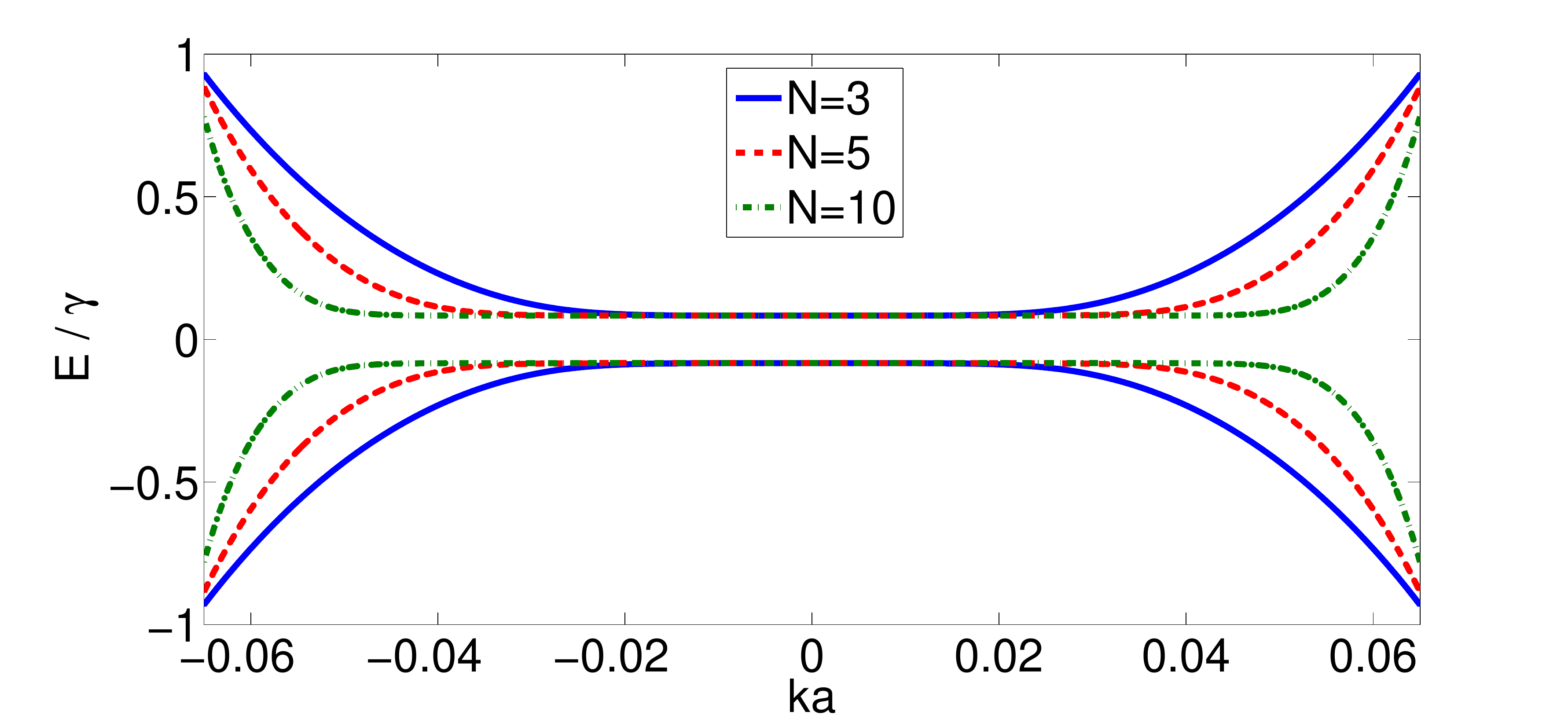}}
\caption{\label{Fig1}  Low energy band structure for $N$ layer ABC stacked graphene in the presence of a vertical electric field. The band structure is plotted in the vicinity of the $\vec{K}$ point, assuming that the potential difference between the top and bottom layers $\Delta = 0.167 t_0 \approx 50 meV$. }
\end{figure}

Now consider the application of a vertical electric field leading to potential difference V between any 2 consecutive layers. We assume that $\gamma \gg e(N-1)V$. 
This ensures that the surface bands are much below the bulk bands. The low energy effective Hamiltonian then takes the form
\begin{equation}
\label{eq6}
H_K(\vec{p}) = \left( \begin{array}{cc} \frac{\Delta}{2} &  \frac{v^N}{(-\gamma)^{N-1}} p_+^N \\  \frac{v^N}{(-\gamma)^{N-1}} p_-^N & -\frac{\Delta}{2} \end{array} \right)
\end{equation}
The resulting band structure is plotted in Fig.\ref{Fig1}. We can see that a gap of magnitude $\Delta=(N-1)eV$ opens in the surface state spectrum. We can also see that the band near the $K$ point is extremely flat, asymptoting to perfect flatness in the limit $N\rightarrow \infty$. 

 \emph{Effect of BN substrate:}  BN is a popular substrate for graphene. It has the same hexagonal structure, but with a slight lattice-mismatch (lattice constant $\sim \frac{56}{55}a$) and is a large band-gap insulator ($E_{gap} \sim 5eV$). 
Placing multilayer graphene on BN substrate introduces a superlattice potential with a $Z \times Z$ supercell, where $Z=56$. The primitive lattice vectors corresponding to the reciprocal lattice are
\begin{equation}
\begin{pmatrix}
\vec{B_{1}} \\
\vec{B_{2}} \\
\end{pmatrix}
=
\frac{2\pi}{3Za}
\begin{pmatrix}
1 & \sqrt{3} \\
1 & -\sqrt{3} \\
\end{pmatrix}
\begin{pmatrix}
\vec{\hat{x}} \\
\vec{\hat{y}} \\
\end{pmatrix}\label{eq: b12}
\end{equation}
The strong superlattice potential is seen primarily by states on the lower surface of the multilayer graphene, close to the substrate. These states are either conduction band or valence band states, depending on the sign of the vertical electric field. We assume the electric field is chosen such that it is the conduction band that mainly sees the superlattice potential. If the multilayer graphene were sandwiched between BN sheets, then both conduction and valence bands would see a strong superlattice potential, irrespective of the sign of electric field. 

The reduced `Brillouin zone' for the superlattice is hexagonal, but scaled by a factor $1/Z$ with respect to the original Brillouin zone. If Z is chosen such that $|\vec{K}|/Z < \gamma/v$, i.e. $Z>Z_c$, where 
\begin{equation}
Z_c = 20\pi/\sqrt{3}=36.3 \label{eq: Zc}
\end{equation}
 then the `flat pockets' at $K$ and $K'$ extend over the entire reduced Brillouin zone. Moreover, umklapp scattering at the boundaries of the reduced Brillouin zone opens a gap between the flat pocket and the rest of the band. Since for BN substrate, $Z = 56 > Z_c$, it follows that placing chiral multilayer graphene on BN leads to flat bands that extend over the entire reduced Brillouin zone, and which are separated from the rest of the bandstructure by the umklapp energy scale $\lambda$. 


\begin{figure}
\includegraphics[width = \columnwidth]{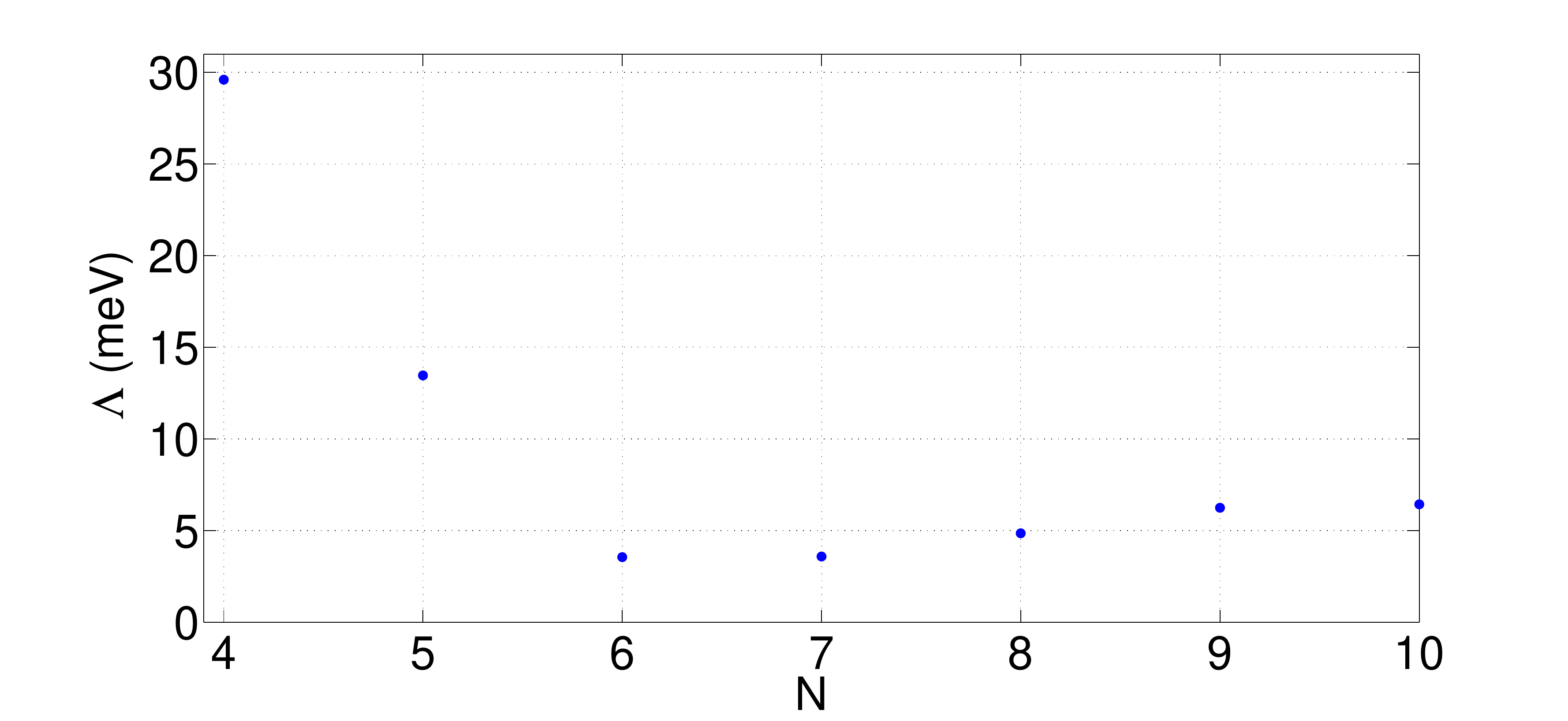}
\caption{\label{fig4} Bandwidth $\Lambda$  of the lowest conduction band for $N$ layer chiral graphene. For $N>5$, the bandwidth comes mainly from umklapp scattering at the zone boundary.}
\end{figure}
 
We now explicitly calculate the bandstructure in the reduced zone. It may be readily determined that for the above superlattice potential, the $K$ and $K'$ points in the original zone get folded to inequivalent corners $\tilde K$ and $\tilde K'$ of the reduced zone. Meanwhile, the reciprocal lattice vectors of the superlattice satisfy $ |\vec{B}_{1,2}| < \gamma/v < 2 |\vec{B}_{1,2}| $, where $\gamma/v$ is the width of the flat pocket. Since Bragg scattering is only effective between states that are near degenerate, we restrict our attention to Bragg scattering by a single reciprocal lattice vector, and neglect higher order Bragg scattering events. The matrix elements of the superlattice potential  $\langle \vec{k} | \hat V | \vec{k+B_{1,2}} \rangle$ were determined by modelling the BN superlattice as a positive $\delta$-function potential at each B site and a negative $\delta$-function potential at each $N$ site \cite{supplement}. Hence we obtained the bandstructure in the reduced zone.  Although the bandstructure contains $Z^2=56^2 = 3136$ conduction bands per spin and valley, only the lowest conduction band is flat, and is separated from the higher bands by an energy scale of order $\lambda$, where the gap may be estimated from the DFT calculations in \cite{Lichtenstein}, and is of order $10 meV$. Meanwhile, the bandgap $\Delta$ between conduction and valence bands can be externally controlled using gates, and may be made as large as desired. We should take $\Delta \gg 10 meV$ to ensure that we do not mix conduction and valence bands. For specificity, we suggest using $\Delta = 50 meV$, which is easily achievable by gating \cite{Crommie}.

The superlattice potential from the BN substrate also introduces intervalley tunneling of magnitude $\lambda/Z$ \cite{supplement}, which turns the bandcrossings of the two nearly flat bands coming from $K$ and $K'$ valleys into avoided crossings. The resulting bandstructure contains two strictly non-degenerate flat conduction bands, separated by a minigap $\lambda/Z\approx 0.2 meV$. The lowest energy band comes mostly from the $K$ valley in regions closest to the $\tilde K$ corner of the reduced zone, and comes mostly from the $K'$ valley in regions closest to the $\tilde K'$ corner. From the bandstructure calculations, we can extract the bandwidth of the two low-lying conduction bands. 
In Fig.\ref{fig4}, we plot the bandwidth of the lowest conduction band as a function of $N$. We see that for $N>5$ the two low-lying conduction bands are nearly flat, with a small residual bandwidth around $5$ meV which comes mainly from umklapp scattering at the zone boundary. The $N=7$ layer system actually has minimum bandwidth of $\sim 3.6$ meV, due to a cancellation between `intrinsic' curvature of the band and the effect of umklapp scattering. 

We also note that there is an indirect band gap separating the two flat bands from the higher energy non-flat bands. It vanishes for $N<6$ and then decreases with increasing $N$. Transitions across the indirect band-gap must be phonon assisted and should be weak, but nevertheless it is essential that we work with $N>5$ to have a truly isolated flat band. Fortunately, $N=7$ is optimal not only through having the flattest band (Fig.\ref{fig4}), but also because its indirect band gap is of order the direct band gap. For $N=7$ the direct band gap is $\sim 6 meV$.

%

We have now verified that the lowest conduction band can be made flat. Now we show that this band has non-vanishing Berry curvature. The momentum-space Berry curvature for the $n$th band is given by \cite{TKNN}
\begin{equation}
B_{n}(\vec{k}) = - \sum_{n' \neq n} \frac{2 Im\left\langle n \vec{k} | v_x | n' \vec{k} \right\rangle \left\langle n' \vec{k} | v_y | n \vec{k} \right\rangle}{(E_{n'}-E_{n})^2}
\end{equation}
where $v_{x(y)}$ is the velocity operator and $E_n$ is the eigenenergy corresponding to the $\left | n \vec{k} \right\rangle$ eigenstate. It may be readily determined that the $K$ and $K'$ bands have opposite signs of local Berry curvature. The conduction and valence bands also have opposite signs of local Berry curvature at every point in momentum space. As a result, when the chemical potential is placed between conduction and valence bands, the system displays quantum valley Hall effect \cite{NandkishoreQAH}. However, flat band physics will manifest itself when the chemical potential is placed inside either the conduction or the valence bands. In this case we can focus  on the band that contains the chemical potential. 

\begin{figure}
\includegraphics[width = \columnwidth]{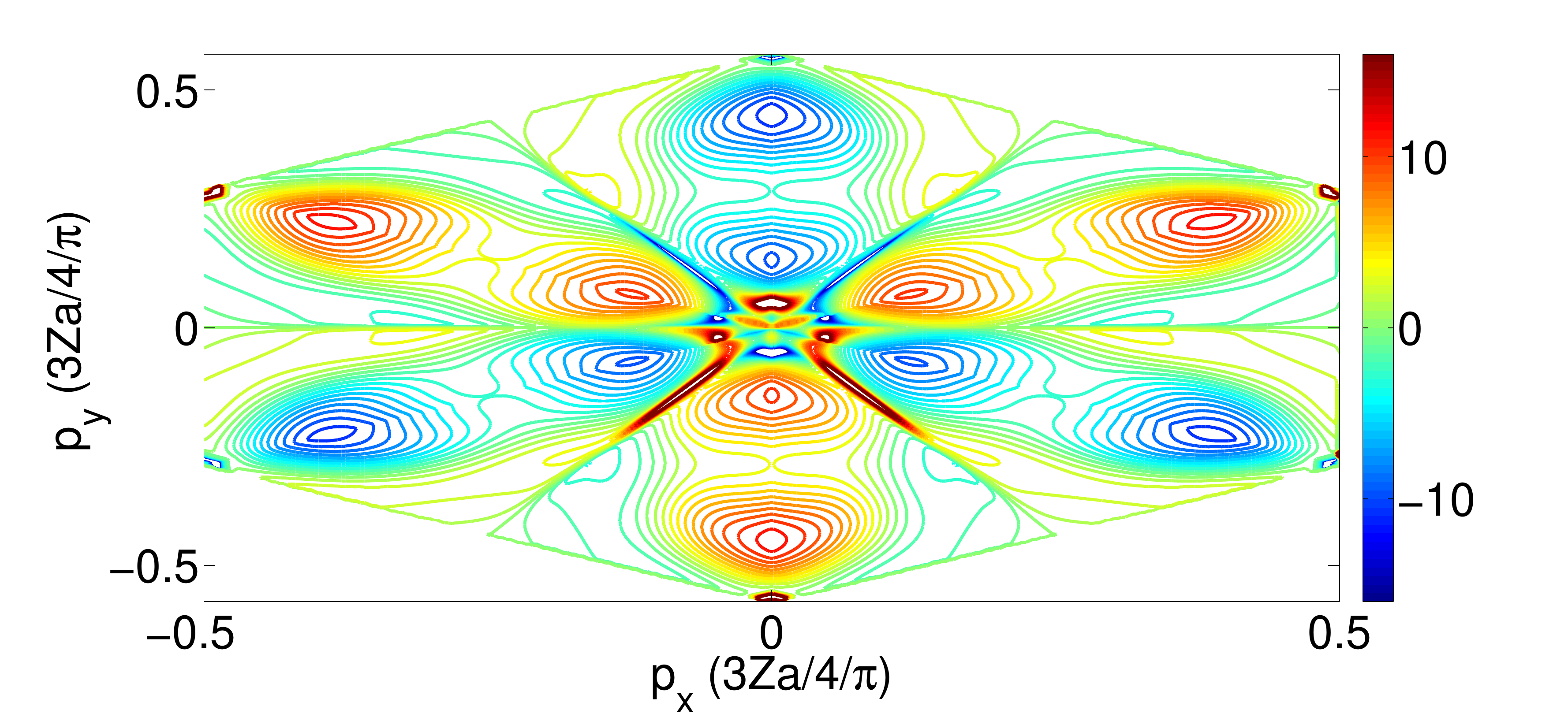}
\caption{\label{fig: curvature} Contour plot of Berry curvature in the lowest flat conduction band for $N=7$. The red/yellow regions come mainly from the $K$ valley and have positive curvature, while the blue regions come mainly from the $K'$ valley and have negative curvature. The Berry curvature integrated over the band is zero.}
\end{figure}

We consider the conduction band for specificity. Due to avoided crossings between bands coming from the two valleys, the lowest conduction band contains regions with positive and negative local Berry curvature respectively while the integrated Berry curvature (Chern number) is zero (Fig.\ref{fig: curvature}). Meanwhile, there is a second nearly flat conduction band which is separated by a mini gap equal to the inter valley scattering amplitude. Interactions may enhance this minigap, lifting the near degeneracy of the two conduction bands to gain exchange energy, in accordance with  `quantum Hall ferromagnetism' \cite{Girvin}. 

Thus, we conclude that placing $N=7$ layer ABC graphene on BN substrate allows us to realize two nearly flat bands with local Berry curvature but zero Chern number which are separated from the non-flat bands by a band gap of $6 meV$ and are separated from each other by a minigap of $0.2 meV$, equal to the inter valley scattering amplitude.  Chiral multilayer graphene thus offers a promising playground for investigating the effect of strong interactions in the presence of a non-trivial quantum distance metric. 

\emph{Chern number from adatoms}. The key feature of chiral multilayer graphene is that it allows access to a system with a non-trivial quantum geometry but without Chern number. However, a non-zero Chern number may also be obtained by making use of adatom deposition on the outermost graphene layers to open up a gap between conduction and valence bands, rather than using vertical electric field. Adatom deposition also introduces a superlattice potential. For the appropriate choice of (time reversal symmetry breaking) adatoms, the two valleys acquire the same sign of Berry curvature \cite{UTA, German group, Chinese group}. The rest of the analysis proceeds exactly as before, only now the Berry curvature has the same sign everywhere in the flat band, and thus does not cancel. 

\emph{Concluding remarks}. We have shown that ABC stacked multilayer graphene placed on BN substrate has a bandstructure containing flat bands. The seven layer system is ideal for this purpose. The flat bands have nonzero local Berry curvature but zero Chern number. Thus, chiral multilayer graphene represents an exciting new frontier in the study of interaction effects in systems with non-trivial quantum geometry, allowing access to an interaction dominated system with a non-trivial quantum distance metric but without the complication of a non-zero Chern number. 

\emph{Acknowledgement} We thank S.L.Sondhi for extensive discussions. We acknowledge useful conversations with Rahul Roy, E.J.Mele, L.S. Levitov, B.A.Bernevig, and L. Santos. A.K. is supported by NSF grant DMR-1006608.

\begin{widetext}

\section{Supplement}

\textbf{A.} We consider the case of multilayer graphene with interlayer hopping $\gamma \neq 0$. It is convenient to separate the low energy and high energy subspaces by writing the Hamiltonian in the $(\psi_{1A\vec{k}} \psi_{NB\vec{k}} \psi_{1B\vec{k}} \psi_{2A\vec{k}} \psi_{2B\vec{k}} \psi_{3A\vec{k}} \cdots \cdots \psi_{(N-1)B\vec{k}} \psi_{NA\vec{k}})$ basis. The Schrodinger equation in this basis becomes
\begin{equation} 
\begin{pmatrix}
H_{11} & H_{12} \\
H_{21} & H_{22} \\
\end{pmatrix}
\begin{pmatrix}
\psi_{low} \\
\psi_{high} \\
\end{pmatrix}
=
\epsilon
\begin{pmatrix}
\psi_{low} \\
\psi_{high} \\
\end{pmatrix}.
\label{eq: H1}
\end{equation}
Here $\epsilon$ is the energy eigenvalue, $\psi_{low}$ is a two component spinor, $\psi_{high}$ is a $2N-2$ component spinor, $H_{11}$ is a 2x2 zero matrix and $H_{12}$ and $H_{22}$ are 2 $\times$(2N-2) and (2N-2)$\times$(2N-2) matrices respectively. Here $H_{12}$ and $H_{22}$ have the form below
\begin{equation} 
H_{12} = 
\begin{pmatrix}
v\vec{p}_{+} & 0 & \cdots & 0 & 0 \\
0 & 0 & \cdots & 0 & v\vec{p}_{-} \\
\end{pmatrix}
\end{equation}
\begin{equation}
H_{22} = 
\begin{pmatrix}
0 & \gamma & 0 & 0 & 0 & \cdots \\
\gamma & 0 & v\vec{p}_{+} & 0 & 0 & \cdots \\
0 & v\vec{p}_{-} & 0 & \gamma & 0 & \cdots \\
0 & 0 & \gamma & 0 & v\vec{p}_{+} & \cdots \\
0 & 0 & 0 & v\vec{p}_{-} & 0 & \cdots \\
\vdots & \vdots & \vdots & \vdots & \vdots & \ddots \\
\end{pmatrix}
\end{equation}
We can eliminate the high energy dimer components in the system of equations (\ref{eq: H1}) to get a Schrodinger equation for only the low energy components in the manner of \cite{McCann}. Since we are interested in only the low energy states, we make a binomial expansion to the first order in $\epsilon$ assuming it is very small compared to $H_{22}$. We get $(H_{eff}-\epsilon)\psi_{low}=0$ where $H_{eff}=(1+H_{12}(H_{22})^{-2}H_{21})^{-1}(H_{11}-H_{12}(H_{22})^{-1}H_{21})$. This leads to a low energy single particle Hamiltonian for the surface states of an $N$ layer graphene which takes the form
\begin{equation}
H_K(\vec{p}) = \frac{v^N}{(-\gamma)^{N-1}} \left( \begin{array}{cc} 0 & p_+^N \\ p_-^N & 0 \end{array} \right); \quad p_{\pm} = p_x \pm i p_y
\end{equation}

\textbf{B.} First, we calculate the matrix elements of the external potential $\hat V$ coming from the BN between Bloch states, $\langle \vec{k} | \hat V | k + \vec{q} \rangle$. This is just equivalent to calculating the Fourier transform of the BN potential. We model the BN potential as $\lambda' \delta(\vec{r_i})  - \lambda' \delta(\vec{r_j})$, where $\vec{r_i}$ are the positions of boron atoms and $\vec{r_j}$ are the positions of nitrogen atoms. This is a natural model, since the atomic numbers of $B$ and $N$ are one less than and one more than carbon respectively. Taking the delta functions to have slightly different weights will not qualitatively alter our results. The boron atoms sit on the A sublattice of the hexagonal superlattice, and the nitrogen atoms sit on the B sublattice. 

The Fourier transform of the above potential takes the form 
\begin{equation}
\langle \vec{k} | \hat V | \vec{k} + \vec{q} \rangle = \hat V(\vec{q}) = \delta_{\vec{k}, \vec{Q}} f_{1}(\vec{q})
\end{equation}
where $\vec{Q}$ denotes a reciprocal lattice vector of the BN lattice (which is equal to $\vec{B_{1,2}}$ in Eq.\ref{eq: b12} of the main text modulo reciprocal lattice vectors of the graphene lattice) and $f_1(\vec{q})$ is a form factor coming from the two site nature of the BN unit cell. For the model potential under consideration, $f_1(\vec{q}) \propto (e^{i \vec{q}.\vec{r_i}} - e^{i \vec{q}.\vec{r_j}})$. Hence, we obtain $\langle \vec{k} | \hat V | \vec{k} + \vec{B_1} \rangle = i \lambda$, $\langle \vec{k} | \hat V | \vec{k} + \vec{B_2} \rangle = i \lambda$, $\langle \vec{k} | \hat V | \vec{k} + \vec{B_1}+\vec{B_2} \rangle = i \lambda$, and $f_1(-\vec{k}) = f^*_1(\vec{k})$. 

Thus far we have assumed that the BN superlattice potential can be modelled as a delta function array. In fact, the B and N atoms carry $+$ and $-$ charge respectively. The BN superlattice potential may thus be better modelled as a delta function array convolved with a $1/r$ envelope. The $1/r$ envelope simply reflects the Coulomb potential arising from a local charge imbalance. This Fourier transforms to a delta function array multiplied by a $1/k$ envelope. Thus intervalley scattering by $Z$ reciprocal lattice vectors is weaker than intravalley scattering by one reciprocal lattice vector by a factor of $1/Z$. However as we will see below, this weak intervalley scattering is still important along lines in the reduced zone where the $K$ and $K'$ bands are degenerate. 

Now we calculate the band structure. First we consider the low energy hamiltonian of multilayer graphene without BN substrate in Eq.\ref{eq6}. It is written in a basis such that (1, 0) is a Bloch state in the A sublattice of the top layer and (0, 1) is a Bloch state in the B sublattice of the bottom layer. Now we consider the case of multilayer graphene sandwiched between BN sheets. Without considering the folding of the bands at the reduced zone edges, the 4 low energy bands corresponding to a particular hexagonal unit cell with center $\vec{Q}$ are found from the eigenvalues of   
\begin{equation}
m(\vec{k},\vec{Q}) =
\begin{pmatrix}
\Delta/2 &\frac{v^N((k_x-Q'_{x})-i((k_y-Q'_{y}))^N}{(-\gamma)^(N-1)} & \frac{\lambda}{Z} & 0 \\
 \frac{v^N((k_x-Q'_{x})+i((k_y-Q'_{y}))^N}{(-\gamma)^(N-1)} & -\Delta/2 & 0 & \frac{\lambda}{Z} \\
\frac{\lambda}{Z} & 0 & \Delta/2 &\frac{v^N(-(k_x-Q''_{x})-i((k_y-Q''_{y}))^N}{(-\gamma)^(N-1)} \\
0 & \frac{\lambda}{Z} & \frac{v^N(-(k_x-Q''_{x})+i((k_y-Q''_{y}))^N}{(-\gamma)^(N-1)} & -\Delta/2 \\
\end{pmatrix}
\end{equation}
where $\vec{Q'}$ points to the $K$ point of the hexagon, for which the distance between $\vec{k}$ and $\vec{Q'}$ is the minimum. A similar definition holds for $\vec{Q''}$ and $K'$ point. The matrix is written in a basis where $(1,0,0,0)$ denotes a state on sub lattice A and with valley $K$, (0,1,0,0) denotes a state on sub lattice $B$ and valley $K$, $(0,0,1,0)$ denotes a state on sub lattice A and valley $K'$ and $(0,0,0,1)$ denotes a state on sub lattice B and valley $K'$. In the off diagonal blocks we have included a weak inter-valley scattering coming from the superlattice potential. 

 The band folding at the zone edges can be taken into account by considering the four low energy bands arising from a central hexagonal unit cell in the reciprocal space and those arising from it's 6 nearest neighbouring cells. The corresponding band structure is easily found from the eigenvalues of the following $7\times7$ matrix:
\begin{equation}
\begin{pmatrix}
m(\vec{k},\vec{K_{0}}) & n^{*} & n & n & n & n^{*} & n^{*} \\
n & m(\vec{k},\vec{K_{1}}) & n & 0 & 0 & 0 & n^{*} \\
n^{*} & n^{*} & m(\vec{k},\vec{K_{2}}) & n & 0 & 0 & 0 \\
n^{*} & 0 & n^{*} & m(\vec{k},\vec{K_{3}}) & n^{*} & 0 & 0 \\
n^{*} & 0 & 0 & n & m(\vec{k},\vec{K_{4}}) & n^{*} & 0 \\
n & 0 & 0 & 0 & n & m(\vec{k},\vec{K_{5}}) & n^{*} \\
n & n & 0 & 0 & 0 & n & m(\vec{k},\vec{K_{6}}) \\
\end{pmatrix}
\end{equation}
where $\vec{K_{0}}$ points to the center of the central hexagon, $\vec{K_{1}}$, $\vec{K_{2}}$, $\vec{K_{3}}$, $\vec{K_{4}}$, $\vec{K_{5}}$ and $\vec{K_{6}}$ point to the centers of the 6 adjoining hexagons. The matrix is written in a basis where $(1,0,0,0,0,0,0)$ is a state near $\bm{K_0}$, $(0,1,0,0,0,0,0)$ is a state near $\bm {K_1}$, $(0,0,1,0,0,0,0)$ is a state near $\bm{K_2}$ etc.  and the matrix corresponding to the intra-valley scattering between the valleys of neighbouring hexagonal unit cells is
\begin{equation}
n =
\begin{pmatrix}
i\lambda & 0 & 0 & 0 \\
 0 & i\lambda & 0 & 0 \\
0 & 0 & i\lambda & 0 \\
0 & 0 & 0 & i\lambda \\
\end{pmatrix}.
\end{equation}
This $4\times4$ matrix is written in the same basis as Eq.13, where $(1,0,0,0)$ denotes a state on sub lattice A and with valley $K$, (0,1,0,0) denotes a state on sub lattice $B$ and valley $K$, $(0,0,1,0)$ denotes a state on sub lattice A and valley $K'$ and $(0,0,0,1)$ denotes a state on sub lattice B and valley $K'$. 

\end{widetext}

\end{document}